\documentclass[aps,prb,groupedaddress,longbibliography,preprint]{revtex4-1}

\usepackage{graphicx}
\usepackage{dcolumn}
\usepackage{bm}
\usepackage{xcolor}
\usepackage{blindtext}
\usepackage{graphicx}
\usepackage{amssymb}
\usepackage{amsfonts}
\usepackage{amsmath}
\usepackage{textcomp}
\usepackage{xcolor}
\usepackage{ulem}
\usepackage{bm}
\usepackage{parskip,float,graphicx}
\usepackage[caption = false]{subfig}
\usepackage[colorlinks=true,allcolors=blue]{hyperref}
\usepackage{physics}

\begin{document}

\title{Injection of anomalous Hall current: the role of impedance matching}

\author{D. Lacour}
\affiliation{Institut Jean Lamour UMR 7198 CNRS, Universit\'e de Lorraine,  Vandoeuvre les Nancy France}
\author{ M. Hehn}
\affiliation{Institut Jean Lamour UMR 7198 CNRS, Universit\'e de Lorraine,  Vandoeuvre les Nancy France}
\author{Min Xu}
\affiliation{Institut Jean Lamour UMR 7198 CNRS, Universit\'e de Lorraine,  Vandoeuvre les Nancy France}
\author{J.-E. Wegrowe} \email{jean-eric.wegrowe@polytechnique.edu}
\affiliation{Laboratoire des Solides Irradi\'es, Ecole polytechnique, CNRS, CEA, Universit\'e Paris-Saclay, F 91128 PALAISEAU, France}

\date{\today}

\begin{abstract}
The electric-power of the anomalous-Hall current injected into a lateral load circuit is studied. The anomalous-Hall current is generated by a $\mathrm{Co_{75}Gd_{25}}$ ferrimagnetic Hall bar and injected into lateral contacts lithographied at the two edges. The current, the voltage and the power injected in the lateral circuit are studied as a function of the magnetization state, the load resistance $R_l$, and the temperature.  The power efficiency shows a sharp maximum as a function $R_l$, which corresponds to the condition of the resistance matching of the two sub-circuits.  The maximum power efficiency is of the order of the square of anomalous-Hall angle. The observations are in agreement with recent predictions based on a non-equilibrium variational approach.
 \end{abstract}

\pacs{72.25.Mk, 85.75.-d \hfill}
\maketitle 

The search for low power consumption electronic devices is one of the main motivation for the development of spintronics. Indeed, the spins attached to the charge carriers allow a direct manipulation of the magnetization states. Accordingly, the power used is that of the spin degrees-of-freedom, and not directly the electric power. Yet the transport of the spins attached to the charge carriers follows the thermodynamic rules that determines Joule dissipation. In the case of Hall effect (HE), anomalous Hall effect (AHE) \cite{AHE}, or spin-Hall effect (SHE) \cite{SHE_Review}, the Joule dissipation is minimized due to the presence of an effective magnetic field that breaks the time-invariance symmetry at the microscopic scale\cite{Onsager}. The effect of the effective magnetic field on the electric carriers is then taken into account by a typical supplementary ``Hall-like" term in the Ohm's law\cite{JAP1,Madon,Matrix,SHE} (see Eq.(\ref{Ohm}) below).

It is well-known that the force associated with a magnetic field - typically the Lorentz force for the HE - cannot produce mechanical work in vacuum (the Lorentz force is always perpendicular to the trajectory). More generally, due to the Onsager reciprocity relations \cite{Onsager,Supp}, it is often assumed that the Hall-current produced by Hall-like effects is dissipationless\cite{Science04,Sinova,Onose,Meng,X_Jin}. However, it is not necessarily the case: typically, the power associated with the Hall voltage in a perfect Hall bar is indeed null, but the power associated with a Corbino disk \cite{Supp} (all other things being equivalent) dissipates \cite{Benda,Madon,JAP3}. This question is the object of the present study. 

The understanding of the injection of spin-current produced by SHE or AHE has attracted much attention in the context of spin-orbit torque (SOT) effects, or charge-to-spin conversion mechanisms. Indeed, SOT allows a ferromagnetic layer to be reversed by the injection of spin-polarized current from an adjacent non-magnetic Hall bar (for the SHE) or from an adjacent magnetic Hall bar (for the AHE)\cite{Stiles1,Mishra,Stiles2,Wu,SOT_Review1,SOT_Review2,SOT_Review3,Buhrman}. These SOT effects hence render possible the direct manipulation of magnetic memory units at nanometric scales on the basis of the spin degrees-of-freedom only. The spin-to-charge conversion would then be effective with minimum Joule dissipation, or even with a totally dissipationless Hall current\cite{Dissipation}.

The present study focuses on the physical conditions that govern the electric-power injection and the efficiency of the anomalous-Hall current. In the experimental protocol proposed here (see Fig.1), the  load circuit is devoted to measure the amount of Joule power injected from the ferrimagnetic layer due to the AHE, regardless of the spin properties \cite{Rque,Otani}. 

The measurements show that the Joule power consumption is indeed very small: it is of the order of the square of the anomalous Hall angle $\theta_{AH}^2 \sim 10^{-5}$. Furthermore, a sharp maximum of the power-efficiency of the current injection is defined by the impedance matching between the magnetic layer and the load circuit. This observation shows that - like for direct spin-injection into semiconductors \cite{Schmidt,Fert} - the impedance matching could also be a crucial issue for metallic interfaces in the Hall configuration.

\begin{figure} 
   \begin{center}
   \includegraphics[width=6.5cm]{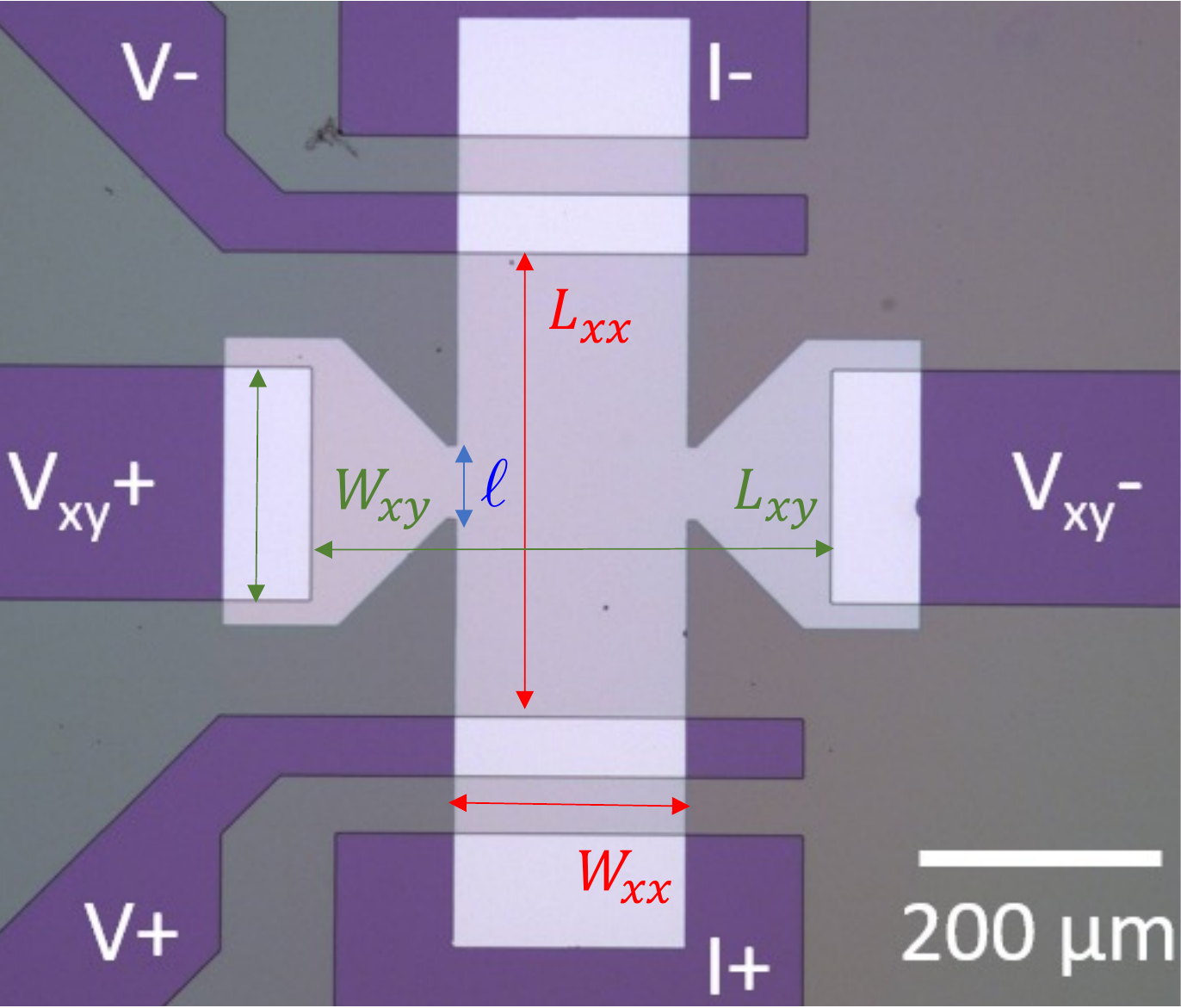} \, \, \, \, \, \, \qquad
     \includegraphics[width=6.5cm]{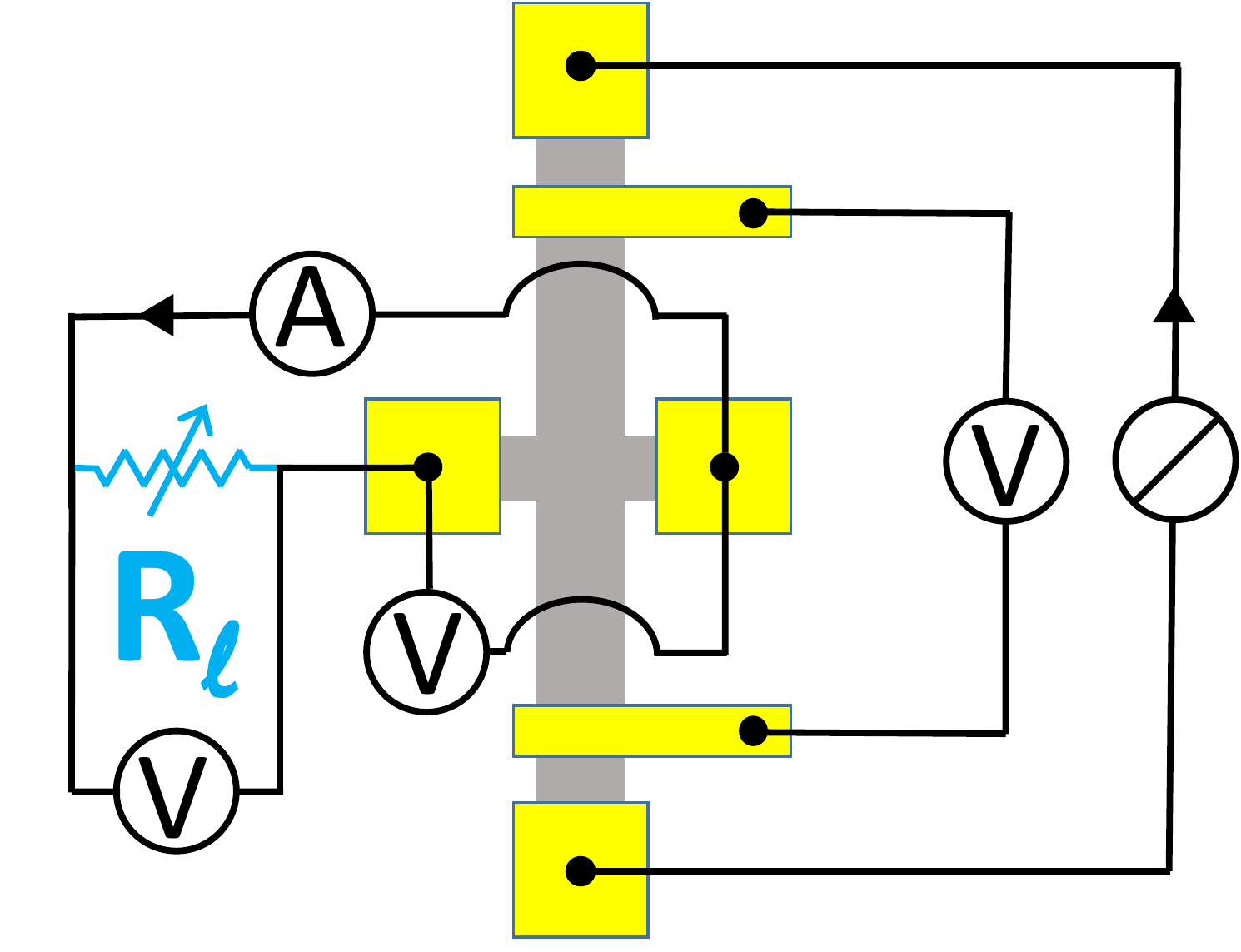}
   \end{center}
   \caption[Schema2]
{ \label{fig1}: (a) Picture of the $CoGd$ Hall bar, with the Au contact pads. The current is injected through the vertical bar (longitudinal current). (b) Circuit with the lateral load resistance $R_l$, the longitudinal-current generator and the position of the ammeter and the voltmeters. }
\end{figure}


The samples is a $\mathrm{Co_{75}Gd_{25}}$  layer of thickness $t=30$ nm, sputtered on a glass substrate and the buffer layer. The choice of the ferrimagnet $\mathrm{Co_{75}Gd_{25}}$ is motivated by its high amplitude of $AHE$, its negligible planar Hall effect (i.e. its negligible anisotropic magnetoresistance) and its negligible magnetocrystalline anisotropy \cite{Madon}.  The magnetic layer is sandwiched between  $5$ nm thick Ta buffer and $3$ nm thick Pt cap. As shown in Fig.1, the length of the Hall bar is $L_{xx} = 400 \, \mu m$ and the width $W_{xx} = 200 \, \mu m$. The magnetic and transport properties of the thin layers have been previously studied (see Supplementary Materials \cite{Supp}). The magnetization is uniform for the quasi-static states under consideration in this study. The out-of-plane shape anisotropy corresponds to a field of about $0.8$ $T$ defined by the magnetization at saturation.  The structure of $\mathrm{CoGd}$ is amorphous and there is no textured.

At room temperature the longitudinal voltage is $\Delta V = 0.148$ $V$ and the resistance of the $CoGd$ layer is $R_{CoGd} \approx 344$ $\Omega$. The anomalous Hall voltage per Tesla is $V_{xy}^0 = 0.886$ $mV$ $T^{-1}$ (the index $^0$ sands for the open circuit). The anomalous Hall angle per Tesla $\theta_{AH} = \frac{ V_{xy}^0}{\Delta V}  = 6 $ $10^{-3}$ $T^{-1}$ is deduced, leading the anomalous resistivity per Tesla of about  $\rho_{AH} \approx 1.02 \, 10^{-2}$ $\mu \Omega.m.T^{-1}$.

Gold contact pads are formed thanks to a standard 2 steps UV lithographic process. As shown in Fig.(\ref{fig1}), two longitudinal pads allow the $\mathrm{Co_{75}Gd_{25}}$ Hall bar to be contacted to the electric generator, and two opposite lateral pads at the edges ($L_{xy} = 450$ $\mu$m and $W_{xy} = 200$ $\mu m$) define the lateral terminals for the load resistances $R_l$, range between $1 \Omega$ and $100 k \Omega $ (decade resistance box). Voltmeters and ampermeter allow the lateral voltage $V_{xy}$ and the lateral current $I_{xy}$ to be measured while  injecting a DC longitudinal current of $0.45$ $mA$ through the $\mathrm{Co_{75}Gd_{25}}$ layer (see Figure 1). An external magnetic field $H_{app}$ varying between $\pm 1.5$ $T$ is applied with from an electromagnet at an angle $\Phi_{\mathrm{Happ}}$ defined with respect to the direction of the injected current $I_{xx}$.

\begin{figure} 
   \begin{center}
  \includegraphics[width=16cm]{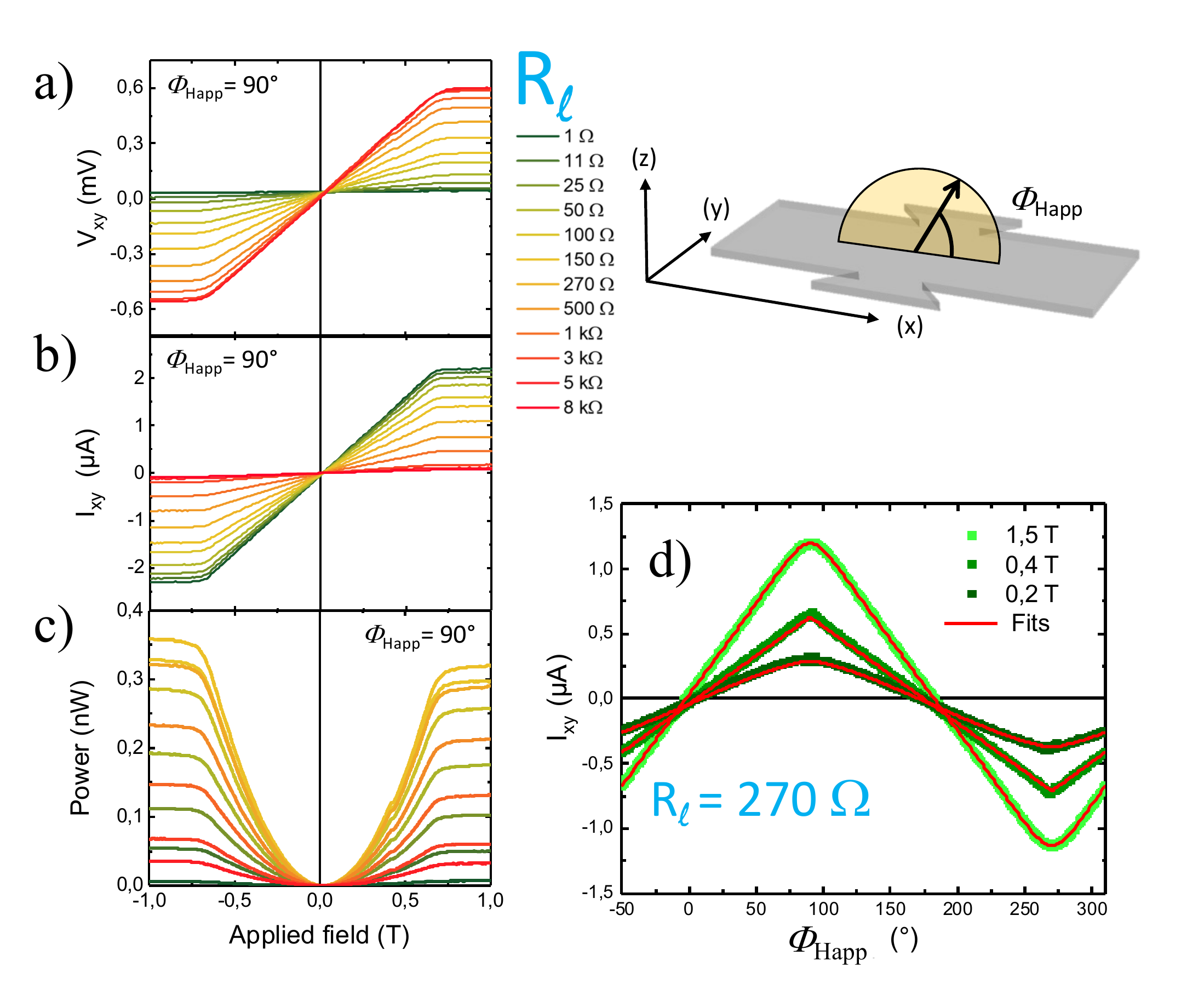}
    \end{center}
   \caption[Schema2]
{ \label{fig2}: (a) Voltage $V_{xy}$ and (b) current $I_{xy}$ measured at the terminals of the load resistance $R_l$ as a function of the magnetic field applied perpendicular to the ferrimagnetic layer. (c) The Joule Power $P= I_{xy} V_{xy}$ measured as a function of the load resistance $R_l$. Current $I_{xy}$ measured as a function of the angle $\Phi_{\mathrm{Happ}}$ of the applied field  at load resistance $R_l = 270 \, \Omega$ for three values of the amplitude of the applied field. The continuous lines are calculated from the AHE. }
\end{figure}

Figure 2 shows the lateral voltage $V_{xy}$ (Fig.2a), the lateral current $I_{xy}$ (Fig.2b) and the power $P = V_{xy} I_{xy}$ (Fig.2c) measured on the load circuit and plotted as a function of the amplitude of the magnetic field at $\Phi_{\mathrm{Happ}}=90^{o}$ (applied normal to the layer) at room temperature. The load resistance $R_l$ is used as a parameter, depicted in the color code shown in the right (green for 1 $\Omega$ up to red for 8000 $\Omega$).

Figure 2d shows the profiles of the anomalous Hall current as a function of the angle $\Phi_{\mathrm{Happ}}$ in a load resistance of $R_l = 270 \, \Omega$ (close to the maximum of the power in the profile of Figure 3b), for three different amplitudes of the applied field.  At $H_{app} = 1.5$ $T$ (at saturation), the magnetization direction $\vec m$ follows approximately the applied field. Below $1$ Tesla, the angle of the magnetization and that of the magnetic field are significantly different due to the shape anisotropy\cite{Supp}.  

The continuous lines in Fig.2d corresponds to the well-known angular profile of the AHE, $V_{xy}(\vec H_{app}) \propto \vec m(\vec H_{app}).\vec e_z $  where $\vec m$ is the magnetization direction and $\vec e_z$ gives the direction normal to the plane of the Hall bar. The contribution of the Planar Hall effect (PHE) of the $CoGd$ layer is about two orders of magnitude below the AHE and can be neglected \cite{Supp}.  The important point is that the observed profiles show a typical signature of the AHE. It is worth pointing-out that the analysis about AHE in Fig.2d is performed on the measurement of the anomalous-Hall {\it current} injected in the Hall circuit, and not on the {\it voltage} of the open circuit: such measurements have not been reported in the literature so far (to the best of our knowledge).

The calculation of $\vec m(\vec H_{app})$ and the details of the magnetic simulation are presented in the supplementary materials \cite{Supp} and in references \cite{Madon}. The excellent agreement between the fit and the data in Fig.2d - together with the specificity of these profiles - confirm that the current measured in the load circuit is due to the anisotropic Hall current generated by the ferrimagnetic Hall bar.

As can be seen in Fig.2a, when a load resistance $R_l$ is contacted to the edges, the amplitude of the anomalous Hall voltage $V_{xy}$ is a monotonous decreasing function of $R_l$. In contrary, the anomalous hall current injected into the lateral circuit is a monotonous increasing function of $R_l$. This is compatible with the intuitive interpretation that the electric charges accumulated at the edges of the Hall bar - as a consequence of the anomalous Hall effect - are extracted and injected into the load circuit in proportion of the load resistance.

 Note that the sign of the anomalous-Hall current is inverted when the direction of the magnetization is changed. This is a consequence of the change of the sign of the accumulated charges when the magnetization is rotated from up to down direction (this sign is equally inverted when direction of the longitudinal current is reversed). The profile of power shown in Fig.2d is well-characterized by a quadratic shape interrupted by the two horizontal lines corresponding to the up and down saturation states of the magnetization \cite{Rque2}.
In Fig.3a the voltage variation (in Volt per Tesla) is plotted as a function of the load resistance $R_l$. 
On the other hand, Fig.3b shows the profile of the Joule power dissipated in the load resistance. The profile of the power is no longer a monotonous function of the load resistance and a sharp maximum appears for a well-defined resistance.  The two typical profiles shown in Fig.3a and Fig.3b are analyzed below. 


\begin{figure}
   \begin{center}
     \includegraphics[width=13cm]{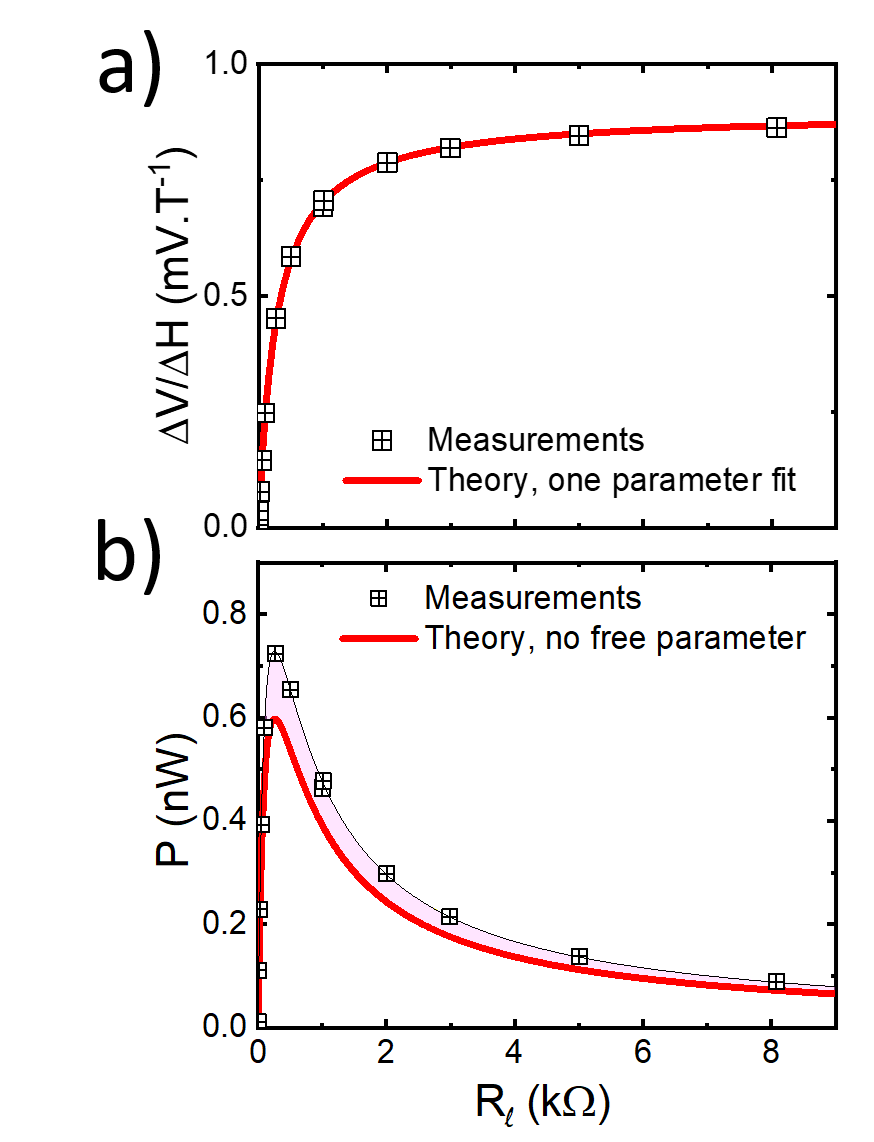}
   \end{center}
   \caption[Py]
{ \label{fig3}: (a) Voltage $V_{xy} $ per unit of magnetic field  plotted as a function of the load resistance $R_l$. The line is a one parameter fit from Eq.(\ref{Voltage}), with the geometric adjustable parameter $\alpha$. (b) Measured electric power $P$ as a function of the load resistance (squares). The lines are the calculation deduced from Eq.(\ref{Power}). The quantitative shift in pink can be attributed to the curvature of the current lines due to the lateral contacts}
\end{figure}

The transverse Hall-current is zero for the open circuit. In contrast, if the Hall bar is in contact to the load circuit, an anomalous-Hall current is generated, that injects the charge carriers accumulated at the edges into the load circuit. The anomalous-Hall voltage decreases accordingly, as seen in Fig.2a.  At stationary regime, the out-of-equilibrium balance between the charge accumulation and current injection into to the load circuit is determined by the minimum power dissipation under the contraints imposed to the system \cite{OnsagerBis,DeGroot}. 

This scenario is investigated in the reference \cite{JAP3} for the usual Hall effect in a perfect Hall bar (assuming the invariance by translation along the longitudinal direction $x$). 

The generalized Ohm's law relates the electric current density $\vec J = \{J_x, J_y, J_z\}$  to the electrochemical potential $\mu$. The gradient $\vec \nabla \mu$ of the electrochemical potential is used instead of the electric field in order to include the effect of the electric screening \cite{JAP3}. The Hall-device is defined in the plane $\{x,y\}$, in which $x$ is the longitudinal direction defined by the current injection from the generator, $y$ is the transverse direction, and $z$ is the direction perpendicular to the plane of the layer. The Ohm's law then reduces to:
\begin{equation}
 \vec{J} = - \sigma \left( \vec{\nabla} \mu - \theta_{AH} \, \vec{e_z} \times \vec{\nabla} \mu \right),
 \label{Ohm}
 \end{equation}
where $\times$ denotes the cross product, $\vec \nabla = \{\partial_x ,\partial_y, \partial_z \}$ is the gradient, $\sigma = 1/\left ( \rho \left ( 1+ \theta_{AH}^2 \right ) \right )$ is the conductivity of the $CoGd$ ferrimagnetic layer \cite{Supp} while $\theta_{AH}$ is the anomalous Hall angle introduced above.
The total electric power dissipated in the device is given by $P= \int \vec J.\vec \nabla \mu \, dv$ (integrated over the volume of the device). After performing the functional minimization of the electric power $P$ under the two constraints of global charge conservation and global current conservation, the Hall voltage at the edges is given by the expression (for constant temperature and small charge accumulation) \cite{JAP3}:
\begin{equation}
  V_{xy}(R_l) =  V_{xy}^0 \, \frac{1}{1 + \frac{\rho}{\alpha \, R_l}},
  \label{Voltage}
  \end{equation}
where $\rho$ is the resistivity of the $CoGd$ layer and the unknown {\it geometrical parameter} $\alpha$ is such that the ratio $\rho/\alpha$ defines the resistance $R$ of the active region of the ferrimagnetic Hall cross.
The fit of the data in Fig.3a with the adjustable parameter $\alpha = 0.93 \, 10^{-8}$ $m$ is given by the red curve.  
 This coefficient $\alpha \approx (\ell  \, t)/W_{xx}$ defines the geometry of the lateral current injection through the {\it active part} of the sample, as shown in Fig.1a. Indeed, the anomalous-Hall current is injected through the effective section $\ell . t$ where $t = 3$ $10^{-8}$ m is the thickness of the sample, $\ell$ is the size of the lateral pad in contact to the Hall bar, and $ W_{xx}= 3 \ell$.

On the other hand, the expression of the power reads\cite{JAP3}:
   \begin{equation}
P(R_l) = P_0\, \theta_{AH}^2 \frac{ \frac{ \rho}{\alpha R_l}}{\left (1+ \frac{ \rho}{\alpha R_l} \right )^2} 
    \label{Power}
  \end{equation}
  where $P_0 \approx 64 \, \mu W$ is the input power injected into the CoGd Hall bar. 
  This value takes into account the correction due to the current flowing through the buffer layer and through the cap layer.
  The profile of the power $P(R_l)$ (Eq.(\ref{Power})) is calculated and plotted in Fig.3b (continuous line), together with the measured power (squares).
 

As can be seen, the measurements are qualitatively in agreement with the model. A sharp maximum is measured for a load resistance equal to the resistance $R_l = \rho/\alpha $. The pink zone between the calculated curve and the experimental points shows a shift between the measurements and the theory. It is surprising to see that the calculated profile is below the measured profile, since the model is based on an optimized ideal Hall-bar (invariant by translation).  We ascribe the shift to the under-estimation of the measured anomalous Hall angle $\theta_{AH}$, which is due to the inhomogeneity of the current lines near the lateral contacts. This effect has been discussed in the context of spin-Hall measurements \cite{Neumann}, and is studied in the supplementary materials \cite{Supp}.

This observation hence validates qualitatively and quantitatively the prediction about the {\it maximum transfer theorem} derived in reference \cite{JAP3}, and the low  power dissipated into the load circuit, of the order of $\theta_{AH}^2$ times the power injected in the Hall bar. These measurements confirm the general assumption that the anomalous-Hall current (like the Hall current) is very sober in terms of power consumption, but  invalidate the claim that it is dissipationless.  \\

Before concluding it is important to point-out that - as shown in Fig.4 - the same measurements performed at different temperatures, from  $T=30 $ $K$ to $T= 295$ $K$, does not change the profiles (described by Eqs.(\ref{Voltage}) and Eq.(\ref{Power})). There is no qualitative change in the dissipation regime for the temperature range under consideration.
In our context, suffice it to say that the temperature dependence of the power is defined by the temperature dependence of the parameters $\theta_{AH}$ and  $\rho$ in Eq.(\ref{Power}), whatever the mechanism responsible for the presence of the effective magnetic field : either spin-orbite coupling or Berry curvature (see the discussion of references \cite{Luttinger, Kondo, Berger, Nozieres,Bruno,Haldane}).

 \begin{figure}
   \begin{center}
     \includegraphics[width=10cm]{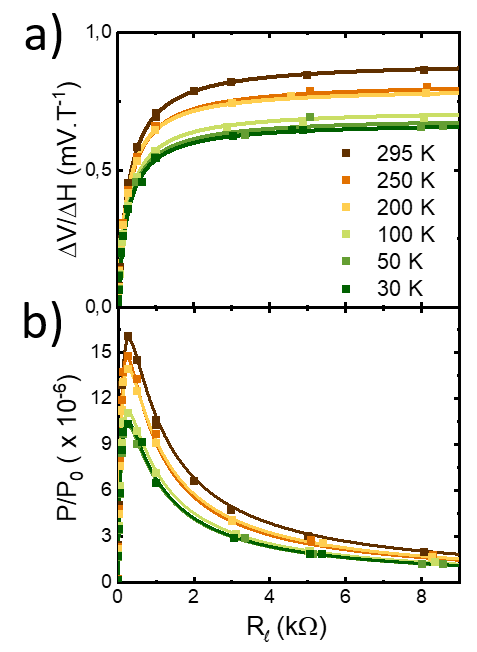}
   \end{center}
   \caption[Py]
{ \label{fig4}:  (a) Voltage $V_{xy}$ per unit of magnetic field  and (b) Normalized electric power $P/P_0$ as a function of the load resistance $R_l$ for six different temperatures. }
\end{figure}

 In conclusion, the anomalous-Hall current and the electric power carried by this current have been studied as a function of the magnetization states $\Phi_{\mathrm{Happ}}$, the load resistance $R_l$ and the temperature $T$. 
The observations show that the maximum electric-power dissipated in the load circuit is indeed very small, of the order of the square of the anomalous-Hall angle $\theta_{AH}^2$. Furthermore, the profile shows a sharp maximum corresponding the resistance matching. The matching occurs when the resistance of the load circuit $R_l$ equals the resistance $R$ of the {\it active part} of the device, defined by the geometrical parameter $\alpha \approx \ell t / W_{xx}$.
 
Last but not least, the presence of a sharp peak of the power injected could have importante consequences for the optimization of SOT. Indeed, the power carried by the spins in a spin-current is supposed to be controlled by the Joule power carried by the electric charges.  The resistance $R_l$ is suspected to play a crucial role in the efficiency of the spin-injection, which could be drastically reduced in case of impedance mismatch.  This study suggests that for the search of a maximum power efficiency, the ensemble of processes responsible for the  SOT could be separated into an upstream anomalous-Hall transducer that transforms a part of the electric power of the generator into the Hall current (Eq.(\ref{Power})), and a downstream charge-to-spin converter that uses the spin-polarization of the current in order to switch a magnetic layer located at nanoscopic distance from the Hall bar.


\begin{thebibliography}{47}
\bibitem{Rque} The measurement of the spin-dependent properties of the anomalous-Hall current in the present experiments would have necessitate to scale down the load circuit at the nanoscopic scale.
\bibitem{Rque2} The small asymmetry between $\pm 1.5$ $T$ fields is due to small offset voltage and current generated at the outside contacts.
\end{thebibliography}
 \end{document}